# Giant optical anisotropy and visible-frequency epsilon-near-zero in hyperbolic van der Waals MoOCl$_2$

*Georgy Ermolaev[1#], Adilet Toksumakov[1#], Aleksandr Slavich[1#], Anton Minnekhanov[1], Gleb Tselikov[1], Arslan Mazitov[1], Ivan Kruglov[1], Gleb Tikhonowski[1], Mikhail Mironov[1], Ilya Radko[1], Dmitriy Grudinin[1], Andrey Vyshnevyy[1], Zdeněk Sofer[2], Aleksey Arsenin[1], Kostya S. Novoselov[3,4,5\*], and Valentyn Volkov[1\*]*

[1]*Emerging Technologies Research Center, XPANCEO, Internet City, Emmay Tower, Dubai, United Arab Emirates*

[2]*Department of Inorganic Chemistry, University of Chemistry and Technology Prague, Prague 6, Czech Republic*

[3]*National Graphene Institute (NGI), University of Manchester, Manchester, UK*

[4]*Department of Materials Science and Engineering, National University of Singapore, Singapore, Singapore*

[5]*Institute for Functional Intelligent Materials, National University of Singapore, Singapore, Singapore*

[#]These authors contributed equally to this work

\*Correspondence should be addressed to the e-mails: vsv@xpanceo.com and kostya@nus.edu.sg

## Abstract

The realization of extreme optical anisotropy is foundational to nanoscale light manipulation. Van der Waals (vdW) crystal MoOCl$_2$ has emerged as a promising candidate for this quest, hosting hyperbolic plasmon polaritons in the visible and near-infrared wavelengths. However, the fundamental anisotropic dielectric tensor governing this behavior has remained elusive. Here, we resolve this problem by providing the first experimental determination of the full dielectric tensor of hyperbolic vdW MoOCl$_2$. Via spectroscopic ellipsometry, Mueller matrix, and reflectance measurements, we quantify the material's optical duality: a metallic optical response ($\varepsilon_1 < 0$) along the crystallographic *a*-axis and a dielectric response ($\varepsilon_1 > 0$) along the orthogonal directions. This dichotomy drives an epsilon-near-zero (ENZ) condition at ≈512 nm and results in giant in-plane birefringence of $\Delta n \approx 2.2$ for MoOCl$_2$. As a result, our work provides the critical missing experimental parameters for MoOCl$_2$, establishing it as a benchmark hyperbolic and ENZ material.

**Keywords:** giant optical anisotropy, hyperbolic material, ellipsometry, Mueller matrix, epsilon-near-zero.

# Introduction

Harnessing light-matter interactions at the nanoscale enables advancements in optical computing[1], polaritonic visualization[2], quantum information[1,3], biosensing[4,5], and integrated photonics[6,7]. The emergence of van der Waals (vdW) materials has provided a revolutionary platform for achieving this control, enabling the design of heterostructures with tunability at the atomic level[8–11]. Within this diverse materials library[12], vdW crystals that possess giant optical anisotropy are of paramount importance[13–17]. Their intrinsic anisotropy is the critical enabling property for a new generation of miniaturized and on-chip photonic devices, offering mechanisms for polarization control[18], subdiffractional light guiding[14,19], and sensing[20]. While many layered materials exhibit strong out-of-plane optical anisotropy[21,22], arising from the fundamental difference between their in-plane covalent bonds and weak out-of-plane vdW forces[14], the discovery of materials with exceptionally strong in-plane optical anisotropy[13,18,23,24] is far more technologically compelling, yet remains a significant challenge.

An extreme and highly sought-after manifestation of optical anisotropy is found in natural hyperbolic vdW materials[25,26]. These materials are defined by a dielectric tensor whose principal components have opposite signs[27]. For instance, exhibiting a metallic optical response (where the real part of dielectric permittivity $\varepsilon_1 < 0$) along one in-plane optical axis and a dielectric optical response ($\varepsilon_1 > 0$) along an orthogonal axis[28]. This condition, which results in an open, hyperboloidal isofrequency surface, allows the material to support highly confined, directional hyperbolic polaritons. Although many vdW materials, such as hexagonal boron nitride (hBN)[26] and α-phase molybdenum trioxide (α-$MoO_3$)[25], exhibit phonon-polaritonic hyperbolicity, this phenomenon is restricted to the mid-infrared spectrum[29]. Therefore, the discovery of natural vdW materials that host these polaritons in the technologically critical visible and near-infrared spectral ranges is a primary goal to advancing the field.

In this quest, the vdW crystal molybdenum oxydichloride ($MoOCl_2$) has emerged as a leading candidate for visible-range hyperbolicity[30–36]. The remarkable properties of MoOCl2 stem from its distinct electronic structure, where an orbital-selective Peierls phase drives the formation of quasi-one-dimensional (1D) Mo-Mo dimerized chains within the vdW layers[37]. This electronic configuration establishes MoOCl2 as a strongly correlated system, often characterized as "bad metal"[35], exhibiting a highly anisotropic Fermi surface and anomalous transport properties, including colossal magnetoresistance[38]. This 1D nature was theoretically predicted to produce a broadband hyperbolic frequency window initiated by a plasma frequency in the visible range, presenting a rare opportunity to access natural epsilon-near-zero (ENZ)[39,40], and spanning hyperbolicity to the near-infrared wavelengths[33,34]. This potential has been recently confirmed by recent pioneering studies[31,32,35,36] that directly visualized hyperbolic plasmon polaritons (HPPs) in $MoOCl_2$ using advanced nano-imaging techniques, including photoemission electron microscopy[36] and scattering-type near-field optical microscopy[32,35]. These works established that $MoOCl_2$ supports low-loss HPPs, a surprising discovery given its high resistivity[38].

However, while these groundbreaking studies[31,32,35,36] have successfully visualized the consequences of hyperbolicity (the propagating HPPs), a fundamental and comprehensive experimental quantification of the root cause remains missing. The full anisotropic dielectric tensor of $MoOCl_2$ that governs this behavior has, to date, been inferred from polariton[32,35] or electronic[36] dispersions or predicted by density functional theory[33,34], but has not yet been directly or systematically measured using gold-standard optical characterization methods. Interestingly, even a standard polarized Raman spectroscopy of $MoOCl_2$ remains largely unexplored.

Here, we resolve this foundational gap by providing the first experimental determination of the giant optical anisotropy and full dielectric tensor of hyperbolic vdW $MoOCl_2$ from ultraviolet (360 nm) to visible and near-infrared wavelengths up to 1700 nm. By employing rotational spectroscopic ellipsometry and Mueller matrix measurements on $MoOCl_2$ flakes, reinforced by a multi-thickness analysis, we extract the complete dielectric tensor. It unambiguously confirms the material's profound duality: $MoOCl_2$ exhibits a metallic optical response ($\varepsilon_1 < 0$) along the crystallographic $a$-axis and a dielectric optical response ($\varepsilon_1 > 0$) along the crystallographic $b$-axis. This difference results in a record in-plane birefringence, demonstrating a superior optical anisotropy of $MoOCl_2$ when benchmarked against other highly anisotropic materials.

## Results

**Structural anisotropy in $MoOCl_2$**

$MoOCl_2$ is a layered van der Waals (vdW) crystal (Figure 1a), belonging to the emerging family of oxydichlorides with the general formula $XOCl_2$, where X = Mo, Nb, Ta, V, or Os.[35] Figure 1a shows the atomic arrangement of $MoOCl_2$ with the unit cell vectors $a$ = 1.277 nm, $b$ = 0.3759 nm, and $c$ = 0.654 nm.[41] The crystal structure is monoclinic ($α = γ$ = 90° and $β$ = 104.8°), characterized by C2/m space group[41]. The defining characteristic of this structure, and the root cause of the material's electronic and optical properties, is an orbital-selective Peierls distortion[37]. In $MoOCl_2$, this distortion is orbital-selective, meaning it affects different electronic orbitals ($d$-orbitals of Mo)[37]. This electronic instability induces physical dimerization of the molybdenum atoms, forming quasi-1D Mo-Mo dimerized chains within a structural backbone of Mo-O-Mo bonds (Figure 1a). These chains are strongly coupled and extend along the crystallographic $a$-axis (Figure 1a). Conversely, along the crystallographic $b$-axis, the inter-chain coupling is mediated by chlorine atoms (Figure 1a) and is significantly weaker. This profound structural anisotropy dictates two distinct types of electronic states: highly delocalized states along the crystallographic $a$-axis, and more localized states along the crystallographic $b$-axis. This structural anisotropy is so dominant that it is evident even at the macroscopic scale. For example, the inset in Figure 1a displays an optical image of a bulk $MoOCl_2$ crystal with needle-like shapes, which often results in flakes with the long edge aligned with $b$-axis and the short edge aligned with $a$-axis.

To experimentally probe this structural anisotropy, we perform polarized Raman spectroscopy, with the results shown in Figures 1b-c (see Supplementary Note 1 for methods). The measurements are conducted with the incident and scattered light polarization aligned parallel to the in-plane crystallographic axes. The Raman spectra along the crystallographic $a$-axis in Figure 1b demonstrate a clear wavelength-dependent intensity of the phonon modes: the highest intensity is observed for 532 nm, then the Raman signal diminishes for 633 nm, and finally the Raman signal almost disappears for 785 nm. This trend is indicative of a metallic optical response along the crystallographic $a$-axis. As the excitation wavelength increases, the penetration depth of the light decreases due to enhanced metallic screening, thereby diminishing the effective Raman scattering volume and suppressing the signal intensity. In contrast, the Raman spectra along the crystallographic $b$-axis demonstrate a similar response for all excitation wavelengths (532 nm, 633 nm, and 785 nm), indicating a dielectric optical response.

In addition, high optical anisotropy of $MoOCl_2$ manifests in a series of polarized optical reflection images of $MoOCl_2$ flakes exfoliated on a silicon substrate captured at varying rotation angles, from 0° to 180°, relative to the incident polarization (Figure 1d and Supplementary Note 2). Figure 1d shows a dramatic color change of $MoOCl_2$ flakes as a function of rotation angle. When the incident polarization is aligned with the crystallographic $a$-axis, the material's response is governed by its metallic nature. Consequently, the flake exhibits a broadband high reflection, appearing bright and mirror-like with a yellowish color (Figure 1d). However, when the flake is rotated by 90° to align the incident polarization with the crystallographic *b*-axis, the material's optical response becomes dielectric. This results in a different light-matter interaction: reflection is significantly reduced, and the material's thickness becomes the dominant factor owing to Fabry-Perot resonances, leading to richer color variation in Figure 1d.

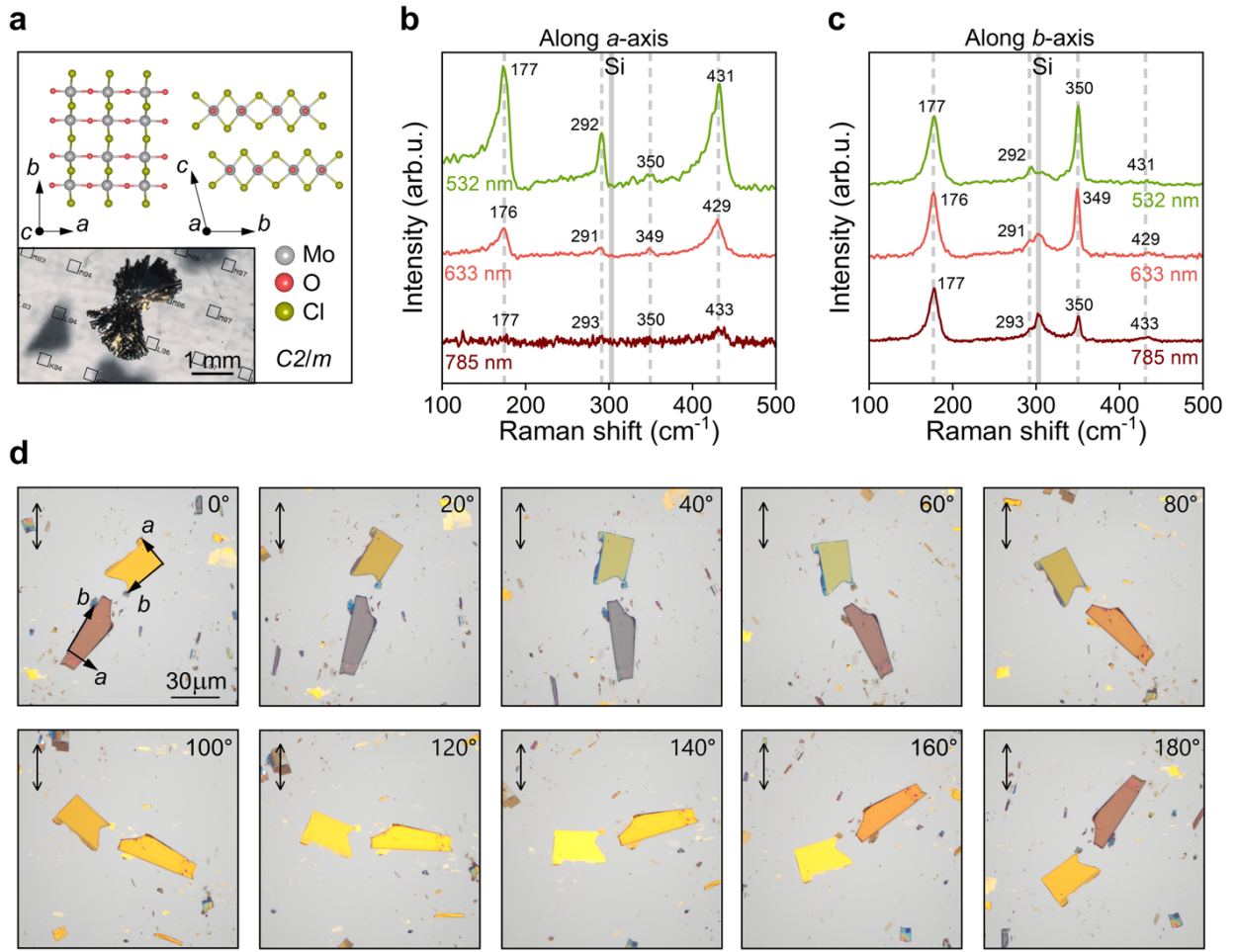

**Figure 1. In-plane structural anisotropy of MoOCl$_2$. a,** Monoclinic crystal structure of MoOCl$_2$. The inset shows an optical image of the bulk crystal. The *z*-direction corresponds to the direction perpendicular to the vdW layers (out-of-plane axis). The polarized Raman spectrum of MoOCl$_2$ along the crystallographic **b,** *a*-axis and **c,** *b*-axis for three excitation wavelengths (532 nm, 633 nm, and 785 nm). The numbers are the peak positions in cm$^{-1}$. Silicon signal is shown by grey region. **d,** Polarized optical reflection images of MoOCl$_2$ flakes on a silicon substrate with varying rotation angle from 0° to 180°. Depending on the rotation angle, flakes demonstrate different colors resulting from the duality of MoOCl$_2$: a metallic ($\varepsilon_1 < 0$) along the crystallographic *a*-axis, but a dielectric ($\varepsilon_1 > 0$) optical response along the crystallographic *b*-axis in the visible range. For more images of MoOCl$_2$ at different rotation angles, see Supplementary Note 2. The polarization direction is shown by arrows in the upper left corner.

**Anisotropic conductivity of MoOCl$_2$**

Anisotropy of MoOCl$_2$ follows also from the direct current (DC) transport measurements on individual flakes. For that, we emply a four-point probe nanoscopy technique (Supplementary Note 1). Figures 2a-b present optical photos of MoOCl$_2$ flakes selected for these measurements. First, we align the four linear probes parallel to the *a*-axis (Figure 2a) to probe the high-conductivity channel and parallel to the *b*-axis (Figure 2b) to probe the low-conductivity channel. To ensure accurate resistivity calculations, the precise thickness of each flake was detrmined via atomic force microscopy (AFM), shown in the insets of Figures 2a-b. The AFM profiles show flat terraces with thickness of $t_a$ = 250 nm for the *a*-axis and $t_b$ = 7510 nm for the *b*-axis. The raw transport data is displayed in Figure 2c as current-voltage (I-V) characteristic curves.

Both crystallographic orientations exhibit linear I-V, confirming the formation of Ohmic contacts. To extract MoOCl$_2$ anisotropic conductivity ($\sigma$) from these measurements, we take into account geometry of the experiment (Supplementary Note 1). As a result, we obtain $\sigma_a \approx 1.35 \cdot 10^6$ S/m and $\sigma_b \approx 1.13 \cdot 10^5$ S/m for the *a*- and *b*-axes, respectively. Comparing these values yields an in-plane conductivity anisotropy ratio of $\sigma_a/\sigma_b \approx 12$. This electronic disparity, reflecting the highly anisotropic Fermi surface, is the fundamental prerequisite for the emergence of optical hyperbolicity driven by anisotropic plasma frequencies.

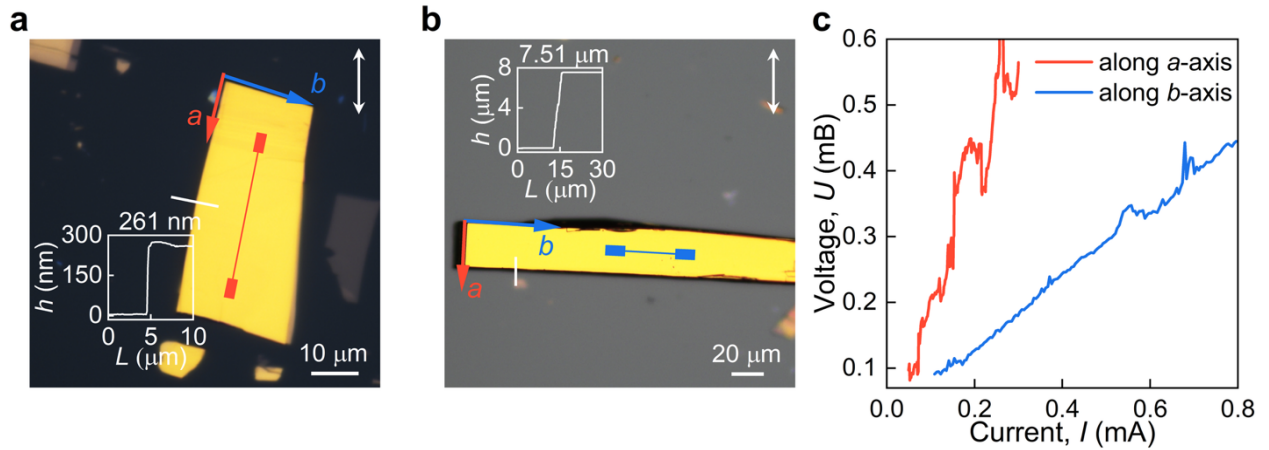

**Figure 2. MoOCl$_2$ anisotropic conductivity measurements.** Optical images of flakes for anisotropic conductivity measurements **a,** along *a*-axis and **b,** along *b*-axis. The lines with rectangles (red for panel (a) and blue for panel (b)) show the place, where four-point probe station was applied. The insets show flakes profiles, measured by atomic-force microscopy. **c,** The current-voltage dependencies for flakes along the corresponding crystallographic axes.

**Spectroscopic ellipsometry of MoOCl$_2$**

To provide the first direct experimental quantification of the full dielectric tensor, we perform variable-angle spectroscopic imaging ellipsometry on MoOCl$_2$ flakes exfoliated on a silicon substrate. We select two flakes with different thicknesses of 2070 ± 70 nm (thick flake) and 115 ± 3 nm (thin flake), as determined by AFM (Supplementary Note 3), to obtain the optical response from both thick and thin flakes. The samples were rotated to align the incident plane of the light parallel to the crystallographic *a*- and *b*-axes. In this configuration, cross-polarization is minimized, allowing the measurement of standard ellipsometric parameters $\Psi$ and $\Delta$ (Figure 3 and Supplementary Note 4) instead of the Mueller matrix. The resulting ellipsometric parameters $\Psi$ and $\Delta$ of the thick flake are presented in Figure 3 (for ellipsometric parameters of the thin flake see Supplementary Note 4). Notably, interference oscillations are evident in the ellipsometry spectra along both axes. Interestingly, both spectra demonstrate interference oscillations, arising from the dielectric cavity along the crystallographic *b*-axis. One might expect these Fabry-Perot fringes only along the crystallographic *b*-axis, but ellipsometry probes the whole dielectric tensor regardless of the flake alignment. Still, the noticeable difference in the measured spectra along the crystallographic *a*-axis (Figure 3a-b) and *b*-axis (Figure 3c-d) already shows a profound optical anisotropy of MoOCl$_2$.

For quantitative analysis, we employ an anisotropic optical model for MoOCl$_2$ with the generic diagonal dielectric tensor diag($\tilde{\varepsilon}_a$, $\tilde{\varepsilon}_b$, $\tilde{\varepsilon}_z$), where $\tilde{\varepsilon}_a$ and $\tilde{\varepsilon}_b$ are the complex dielectric functions for the corresponding crystallographic *a*-axis and *b*-axis, and $\tilde{\varepsilon}_z$ is the complex dielectric function for the direction perpendicular to MoOCl$_2$ vdW layers (see *z*-direction in Figure 1a and Supplementary Note 1). For the initial approximation of the anisotropic optical constants of MoOCl$_2$, we use the theoretically predicted optical constants calculated by Gao and coworkers within the density functional theory (DFT) formalism[33].

To parameterize the anisotropic optical response, we employed a hybrid Drude[42] and Tauc-Lorentz[43] model. The Tauc-Lorentz formalism is particularly well-suited for describing interband transitions in materials with complex electronic structures or localized character, such as many vdW crystals and correlated systems[14,22,44–46], as it properly accounts for the joint density of states near the band edges. As demonstrated in Supplementary Note 5, the Drude-Tauc-Lorentz model provides a superior fit to the experimental data compared to the conventional Drude-Lorentz model[32], capturing the fine spectral features more accurately. As a result, the dielectric function along the crystallographic *a*-axis ($\tilde{\varepsilon}_a = \varepsilon_{a,1} + i\varepsilon_{a,2}$) was modeled with a Drude-Lorentz oscillator term to account for the free-carrier (intraband response) and with the Tauc-Lorentz oscillators to account for the interband transitions. Conversely, the dielectric function along the crystallographic *b*-axis ($\tilde{\varepsilon}_a = \varepsilon_{b,1} + i\varepsilon_{b,2}$) was modeled with a series of the Tauc-Lorentz oscillators without a Drude term. In turn, the dielectric function along the *z*-direction ($\tilde{\varepsilon}_z = \varepsilon_{z,1} + i\varepsilon_{z,2}$) was modeled using the Sellmeier model (see Supplementary Note 5), given that the first-principle calculations[33] suggest a negligible imaginary part ($\varepsilon_{z,2} \approx 0$) for the considered spectral range (360 – 1700 nm). The resulting Drude-Tauc-Lorentz and Sellmeier models parameters are collected in Supplementary Note 5, where we also presented the resulting parameters of the classical Drude-Lorentz model for reference.

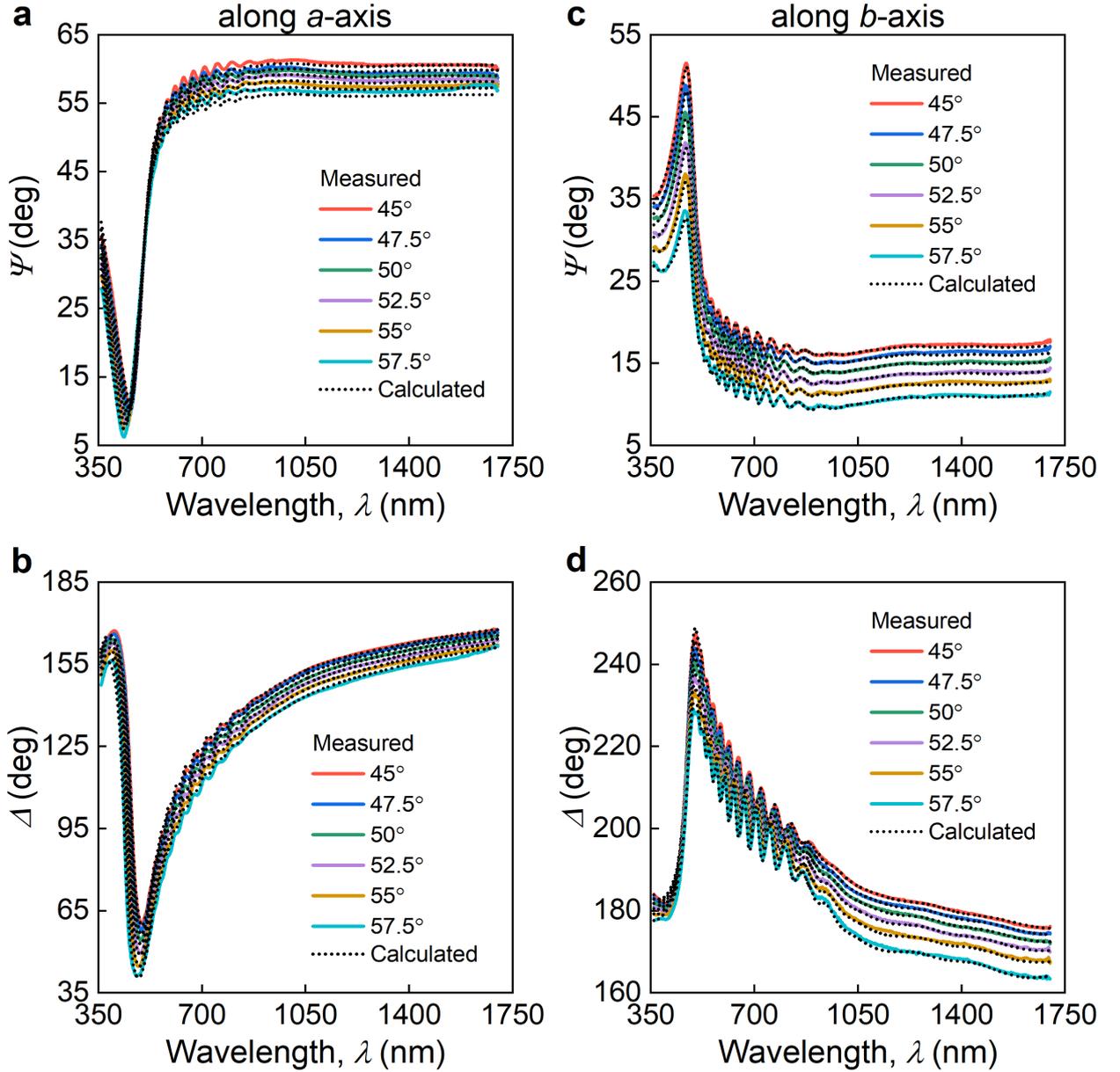

**Figure 3. Ellipsometry spectra of a 2070-nm-thick MoOCl$_2$ flake.** Ellipsometric parameters **a**, $\Psi$ and **b**, $\Delta$ with the incident plane along the crystallographic *a*-axis. Ellipsometric parameters **c**, $\Psi$ and **d**, $\Delta$ with the incident plane along the crystallographic *b*-axis. The measured spectra (solid lines) agree well with the calculated spectra (dotted lines) from the optical model for all the measured incident angles from $\theta$ = 45° to 57.5°.

To validate the fitting of ellipsometry in Figure 3 and Supplementary Notes 4-5, we perform Mueller matrix and reflectance measurements, presented in Figure 4. Unlike ellipsometry, which is aligned along the crystallographic axes in Figure 3, the Mueller matrix under varying in-plane rotation angles $\varphi$ exhibits significant p-to-s and s-to-p polarization conversion. In fact, non-zero off-diagonal Mueller matrix elements ($m_{13}$, $m_{14}$, $m_{23}$, $m_{24}$, $m_{31}$, and $m_{32}$) in Figure 4a are a direct observation of this induced cross-polarization. Therefore, the fact that our optical model of Drude-Tauc-Lorentz/Sellmeier accurately predicts these complex spectra in Figure 4a confirms its physical validity and predictive capability.

An additional validation of our Drude-Tauc-Lorentz/Sellmeier optical model is a polarized normal incidence reflectance measurements in Figures 4b-c. Figure 4b displays the reflectance along the crystallographic a-axis for both 115-nm-thick and 2070-nm-thick flakes. In excellent agreement with our optical model (dotted lines), the measured spectra (solid lines) for both flakes show high (up to 80%) and relatively featureless reflectance in Figure 4b. This behavior is the unambiguous spectral hallmark of the metallic optical response ($\varepsilon_{1,a} < 0$) governed by the strong Drude component and explaining the bright mirror-like appearance observed in polarized optical microscopy (Figure 1d). Conversely, Figure 4c shows the reflectance along the dielectric crystallographic *b*-axis. As expected from our optical model, this reflectance is significantly lower, and thickness-dependent oscillations dominate the spectra. These are classic Fabry-Perot resonances, or interference fringes, arising from the dielectric cavity formed by the transparent flake. The spectral position and periodicity of these fringes are exquisitely sensitive to the optical path length. Therefore, the ability of our optical model to predict the peak-and-valley positions with high precision for both thin and thick flakes (Figure 4c) is a powerful confirmation of our results.

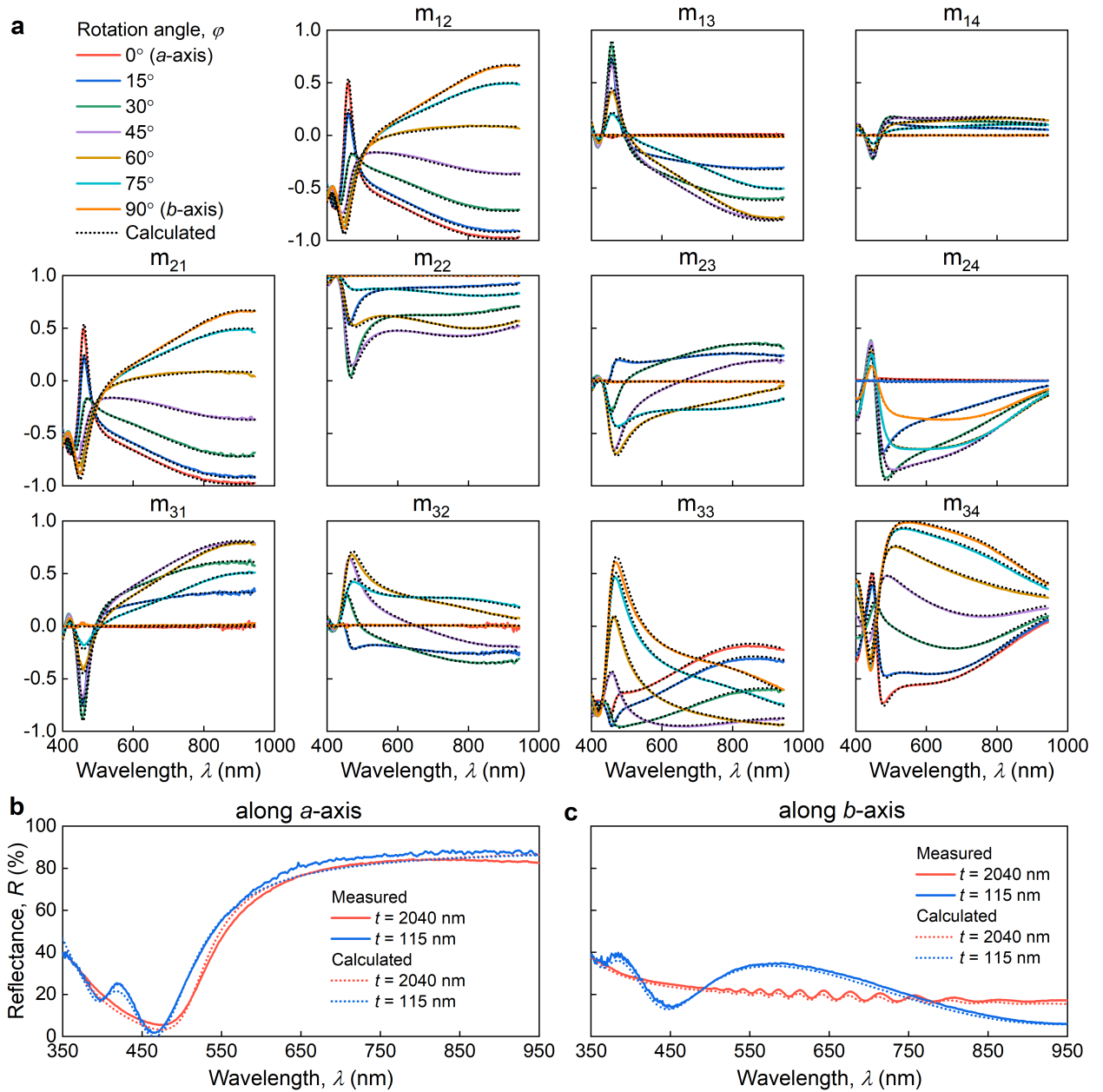

**Figure 4. Mueller matrix and reflectance spectra for MoOCl$_2$ flakes. a,** Mueller matrix elements for 115-nm-thick MoOCl$_2$ flake for different rotation angles $\varphi$ of the flake at $\theta$ = 50° incident angle. The normal incidence reflectance spectra along **b,** *a*-axis and **c,** *b*-axis for two flakes of MoOCl$_2$ with the thicknesses of 2070 nm and 115 nm. The measured spectra (solid lines) agree well with the calculated spectra (dotted lines) for both Mueller matrix elements and reflectance.

The complete anisotropic optical constants of MoOCl$_2$, extracted from ellipsometry fitting (Figure 3 and Supplementary Notes 4-5) and validated by Mueller matrix and reflectance measurements (Figure 4), are presented in Figure 5. This analysis provides the first direct experimental determination of the full dielectric tensor of MoOCl$_2$ and unambiguously confirms the material's hyperbolic properties, that is, metallic-dielectric optical duality. Figure 5a displays the real part ($\varepsilon_1$) of the dielectric permittivity tensor components for the in-plane crystallographic *a*- and *b*-axes, and the out-of-plane *z*-axis, across the ultraviolet to near-infrared spectrum (360 – 1700 nm). As anticipated, along the crystallographic *a*-axis, MoOCl$_2$ exhibits a metallic optical response ($\varepsilon_{1a} < 0$) starting from 512 nm (Figure 5a). We should direct the particular attention to the optical response of the crystallographic *a*-axis around this zero-crossing point because it corresponds to epsilon-near-zero (ENZ) regime. The implications of this transition are twofold. First, the boundary continuity of the electric displacement field ($D_a = \varepsilon_a E_a$) implies a significant enhancement of the internal electric field component ($E_a$) as $\varepsilon_a \to 0$.[47] Second, the dispersive nature of the refractive index near this transition results in a "slow light" regime where the group velocity is significantly reduced. In addition to relatively small optical losses ($\varepsilon_{2a} \approx 0.85$), the anisotropy of MoOCl$_2$ allows for unique mixed-state modes where the ENZ confinement along the *a*-axis can be coupled with dielectric propagation along the *b*-axis, potentially mitigating overall propagation losses for hybrid modes. As a result, unlike noble metals with ENZ in deep ultraviolet[48] or indium tin oxide in the near-infrared[49], MoOCl$_2$ offers accessible ENZ physics in the visible.

At the same time, along the crystallographic *b*-axis and out-of-plane *z*-axis, MoOCl$_2$ shows a purely dielectric behavior ($\varepsilon_{1b,z} > 0$) across the entire measured range. The imaginary part of the dielectric permittivity ($\varepsilon_2$), shown in Figure 5b, quantifies the material's optical losses. The metallic *a*-axis ($\varepsilon_{2a}$) displays the expected Drude absorption tail, where optical losses increase at longer wavelengths, yet it remains relatively low-loss in the visible spectrum. Meanwhile, the dielectric *b*-axis ($\varepsilon_{2a}$) is highly transparent, especially in the visible spectral range (Figure 5b). This low-loss character in all optical axes within the visible spectrum is the fundamental origin of the long-lived hyperbolic plasmon polaritons recently observed in this material[31,32,35,36]. The corresponding anisotropic refractive index (*n*) and extinction coefficient (*k*) are plotted in Figures 5c-d. Of immediate interest is a high in-plane and out-of-plane optical anisotropy of MoOCl$_2$ with in-plane birefringence ($\Delta n$) reaching a record $\Delta n \approx 2.2$ at the maximum (Figure 5e). To contextualize this finding, Figure 5f benchmarks the in-plane optical anisotropy of MoOCl$_2$ with other highly anisotropic materials, including CrSBr[50], As$_2$S$_3$[18], calcite[51], rutile[51], CsPbBr$_3$,[52] GeS$_2$,[53] NbOCl$_2$,[54] Ta$_2$NiSe$_5$,[17] and Ta$_2$NiS$_5$.[16] This comparison demonstrates that MoOCl$_2$ optical anisotropy is superior to the well-known anisotropic systems across the visible and near-infrared wavelengths, confirming its exceptional potential for anisotropic nanophotonics[1,6,7,14,16].

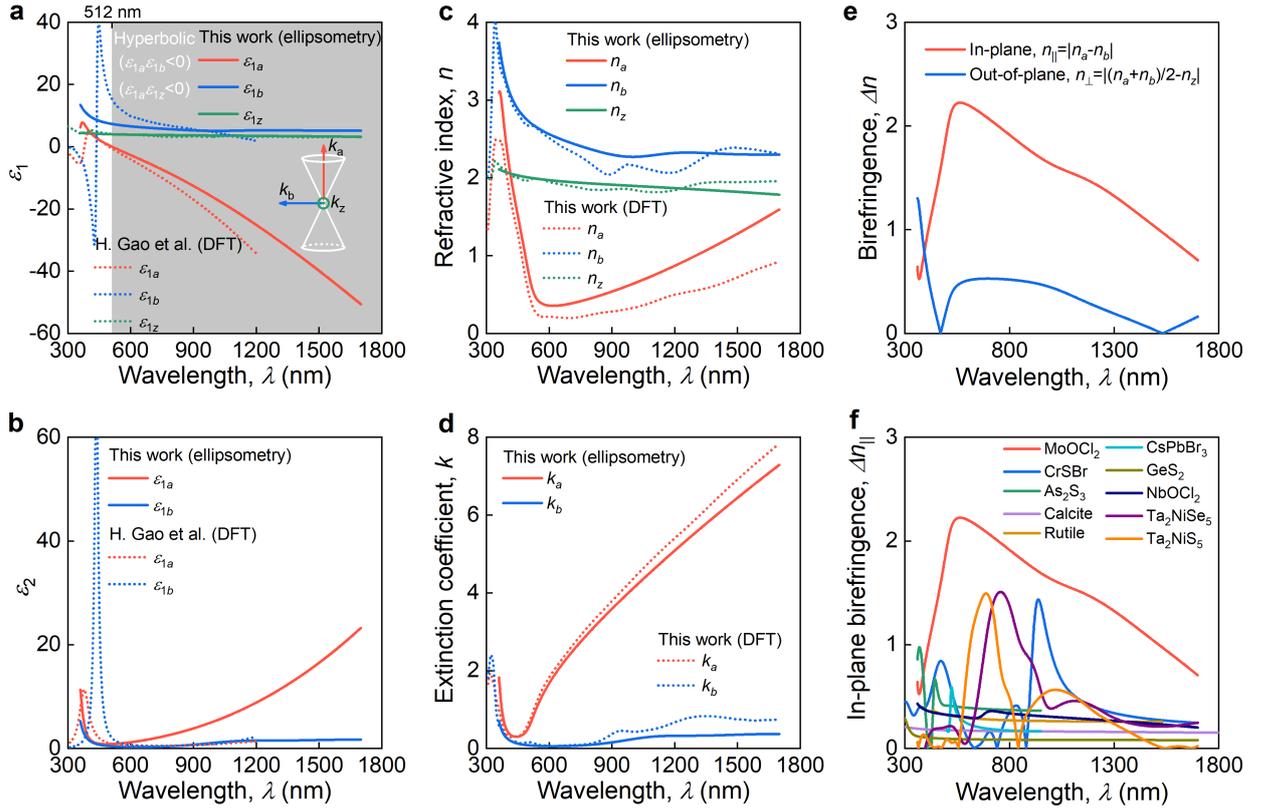

**Figure 5. Anisotropic optical constants of MoOCl$_2$.** The **a,** real and **b,** imaginary parts of the dielectric function of MoOCl$_2$. The gray region shows the hyperbolic window of MoOCl$_2$, where $\varepsilon_{1a}\varepsilon_{1b} < 0$ and $\varepsilon_{1a}\varepsilon_{1c} < 0$ with a hyperboloidal isofrequency surface, right-center inset in panel (a). The boundary of the hyperbolic region is 512 nm, where $\varepsilon_{1a} = 0$. The density functional theory (DFT) optical constants are adopted from the work[33]. The comparison of in-plane optical constants with other experimental works[30,31] reporting MoOCl$_2$ are collected in Supplementary Note 6. Tabulated optical constants of MoOCl$_2$ are collected in Supplementary Note 7, and Drude-Tauc-Lorentz oscillators parameters are collected in Supplementary Note 5. **c,** Refractive index and **d,** extinction coefficient of MoOCl$_2$ from ellipsometry and our GW calculations. **e,** In-plane and out-of-plane birefringence of MoOCl$_2$. **f,** The comparison of in-plane birefringence of MoOCl$_2$ with other highly anisotropic materials. The optical anisotropies for other materials are adopted from several works[16–18,50–54].

## Conclusions and Outlook

In conclusion, we have presented the first experimental determination of the full anisotropic dielectric tensor of MoOCl$_2$, spanning the ultraviolet (360 nm) to the near-infrared (1700 nm) wavelengths. This work resolves a foundational gap in the field, as previous pioneering studies[31,32,35,36] have focused on visualizing the consequences of this material's hyperbolic properties, namely, hyperbolic plasmon polaritons, while the root cause (the dielectric tensor) is unveiled in our study. Using spectroscopic ellipsometry, validated by Mueller matrix analysis and reflectance measurements, we directly observe the profound optical duality of anisotropic optical constants of MoOCl$_2$, which behaves as a metal ($\varepsilon_{1a} < 0$) along the crystallographic *a*-axis and as a dielectric ($\varepsilon_{1b,z} > 0$) along the orthogonal crystallographic *b*-axis and out-of-plane *z*-axis. This extreme difference in the nature of the optical properties results in the highest in-plane optical anisotropy reported for a natural material in the visible and near-infrared spectrum, with a record birefringence reaching values up to 2.2. This giant broadband optical anisotropy establishes MoOCl$_2$ as a compelling platform for next-generation anisotropic nanophotonics. For example, it opens a direct pathway for designing and fabricating ultracompact, on-chip optical components, such as miniaturized polarizers[30], ultrathin waveplates[18], subdiffractional light guiding[19], ultrastrong coupling[55], and Purcell factor enhancement[56]. Furthermore, the experimentally confirmed ENZ transition at 512 nm identifies MoOCl$_2$ as a promising ENZ candidate for nonlinear nanophotonics[57], such as second[58] and third[59] harmonics generations. As a result, the experimentally validated optical constants reported in this work are the critical missing ingredient for the precise design and engineering of functional hyperbolic nanophotonic devices, which can now be reliably modeled and experimentally realized.

## Supplementary Information

Supplementary Information contains sections Materials and Methods, Additional Figures, and tabulated optical constants.

## Author Contributions

G.E., A.T., and A.S. contributed equally to this work. G.E. G.Tselikov, A.A., K.S.N., and V.V. suggested and directed the project. G.E., A.T., A.S., A.Minnekhanov, G.Tselikov, D.G., G. Tikhonowski, M.M. and Z.S. performed the measurements and analyzed the data. G.E., I.R., A.Mazitov, I.K., and A.V. provided theoretical support. G.E. wrote the original manuscript. All authors reviewed and edited the paper. All authors contributed to the discussions and commented on the paper.

## Competing Interests

The authors declare no competing interests.

## Data Availability

The datasets generated during and/or analyzed during the current study are available from the corresponding author upon reasonable request.